\documentclass[onecolumn,trackchanges]{aastex62}

\usepackage{url,hyperref,microtype}
\usepackage{afterpage,longtable}
\usepackage{natbib}
\usepackage{rotating}
\usepackage{subfigure}
\usepackage{multirow}
\usepackage{tabularx,calc}
\usepackage{array} 
\usepackage{url}
\usepackage{epstopdf}
\usepackage{booktabs,fixltx2e}
\usepackage{enumitem}

\usepackage{nccmath}
\usepackage{multirow}
\usepackage{amsmath, amstext}
\usepackage{amssymb}
\usepackage{amsmath}
\usepackage{mathtools}
\usepackage{color} 
\usepackage[flushleft]{threeparttable}
\usepackage{float}
\setlist{leftmargin=0.4in}

\received{March 8, 2016}
\accepted{May 9, 2016}

\shorttitle{BE and Additional BLR Component in AGN with Superluminal Jets}
\shortauthors{Pati\~no-\'Alvarez et al.}

\begin{document}

\title{Baldwin Effect and Additional BLR Component in AGN with Superluminal Jets}

\correspondingauthor{V. M. Pati\~no-\'Alvarez}
\email{victorp@inaoep.mx}

\author{V. M. Pati\~no-\'Alvarez}
\affil{Instituto Nacional de Astrof\'isica, \'Optica y Electr\'onica,  Apartado Postal 51 y 216, 72000 Puebla, Mexico}

\author{J. Torrealba}
\affil{Instituto Nacional de Astrof\'isica, \'Optica y Electr\'onica,  Apartado Postal 51 y 216, 72000 Puebla, Mexico}

\author{V. Chavushyan}
\affil{Instituto Nacional de Astrof\'isica, \'Optica y Electr\'onica,  Apartado Postal 51 y 216, 72000 Puebla, Mexico}

\author{I. Cruz-Gonz\'alez}
\affil{Instituto de Astronom\'ia, Universidad Nacional Aut\'onoma de M\'exico, Ap. 70-264, 04510, DF, Mexico}

\author{T. Arshakian}
\affil{Physikalisches Institut, Universit\"at zu K\"oln, Z\"ulpicher Strasse 77, D-50937 K\"oln, Germany}
\affil{Byurakan Astrophysical Observatory, Aragatsotn prov. 378433, Armenia, and Isaac Newton Institute of Chile, \\ 
Armenian Branch}

\author{J. Le\'on-Tavares}
\affil{Instituto de Astronom\'ia, Universidad Nacional Aut\'onoma de M\'exico, Ap. 70-264, 04510, DF, Mexico}

\author{L.\v{C}. Popovi\'c}
\affil{Astronomical Observatory, Volgina 7, P.O. Box 74 11060 Belgrade, Serbia}

%
%
%
%


\begin{abstract}

We study the Baldwin Effect (BE) in 96 core-jet blazars with optical and ultraviolet spectroscopic data from a radio-loud AGN sample obtained from the MOJAVE 2 cm survey. 
A statistical analysis is presented of the equivalent widths ($W_{\lambda}$) of emission lines H$\beta\,\lambda$4861, 
 Mg II\,$\lambda$2798, C IV\,$\lambda$1549, and continuum luminosities at 5100\,\AA, 3000\,\AA, and 1350\,\AA.  
The BE is found statistically significant (with confidence level \textit{c.l.} $\geq\,$ 95\%) in H$\beta$ and C IV 
emission lines, while for Mg II the trend is slightly less significant (\textit{c.l.} = 94.5\%). The slopes of the BE in the studied samples for H$\beta$ 
and Mg II are found steeper and with statistically significant difference than those of a comparison radio-quiet sample. We present simulations of the 
expected BE slopes produced by the contribution to the total continuum of the non-thermal boosted emission from the relativistic jet, 
and by variability of the continuum components. We find that the slopes of the BE between radio-quiet and radio-loud AGN should not be different, 
under the assumption that the broad line is only being emitted by the canonical broad line region around the black hole. 
We discuss that the BE slope steepening in radio AGN is due to a jet associated broad-line region. 

\end{abstract}

\keywords{galaxies: active --- galaxies:jets  --- quasars: emission lines}


\section{Introduction}

\citet{baldwin77} discovered that quasars follow a relation between the rest frame equivalent widths for the ultraviolet 
lines (e.g., C IV, Ly$\alpha$, etc.) and the continuum luminosity at 1350 \AA\,(L$_{1350}$), known as the
 \textit{Baldwin Effect} \citep{carswell78}, hereafter BE. This relation became quite important and has been the subject 
 of many investigations because it allows to study the physics of the diverse emitting regions present in active galactic
 nuclei (AGN).  The BE is well established for broad emission lines in the ultraviolet and optical regions 
 \citep[e.g.,][]{shields07}, and it is also found that steepens with increasing ionization potential 
 \citep{zheng93, dietrich02}. Most recently, the BE has also been found in narrow emission lines \citep[e.g.,][]{croom02,dietrich02,netzer04,netzer06,netzer07,kov10,popovic11, zhang13}. It has been suggested that the 
 BE could be used to probe the model predictions of the spectral energy distributions (SED) as a function of 
 luminosity \citep{dietrich02}, or to test cosmological models at high redshifts \citep{shields07}.

Despite the advances made in this subject during the last three decades, the physical mechanisms driving the 
observed BE remain unclear \citep[see][and references therein for a complete review on the BE]{shields07}. The 
most widely accepted driving mechanism is that the ionization continuum softens as  the luminosity increases    
\citep[e.g.,][]{zheng93}, so that high-luminosity AGN decrease the fraction of ionizing photons for broad emission 
line formation. This is consistent with \citet{Scott04} that the low-luminosity AGN show harder spectral continuum 
in the extreme-UV. Some theoretical studies support that the  BE is driven,  at least in part, on both the continuum 
shape and the metallicity of the gas  \citep[e.g.,][]{korista98}. 

Other fundamental parameters have been proposed as the principal drivers of the BE: the Eddington ratio
 \citep{baskin04,bachev04,dong09},  or the black hole mass \citep[e.g.,][]{warner03,xu08}. Nevertheless, a 
 consensus on these issues has not yet emerged.

Moreover,  it is now well established that emission lines originated from higher ionization species display steeper 
slopes in the $W_{\lambda}-L_{c}$ diagram. This means that the intensity of the correlation, traced by the slope, 
seems to be dependent  on the  emission line ionizing energy, as was shown by several authors \citep[e.g.,][]{zheng93,zheng95,espey99,dietrich02}.

The aim of this paper is to investigate the BE in radio-loud AGN possessing relativistic jets. This is of great 
interest because, through the past decades, several studies have shown that AGN spectral properties 
differentiate depending on radio-loudness\footnote{Radio-loudness classic criteria $R$:  the ratio between the  
radio (5\,GHz) and optical (4400\,\AA) flux densities  $R\,=\,F\, _{\mathrm{5\,GHz}}/Fo\,_{\mathrm{4400\,A}}$ 
\citep{K89}.}. For example, \citet[][and references therein]{B01} found that the composite spectrum of radio 
loud (RL; $\rm{log}\, \textit{R}\,>\,1$) AGN, compared to that of radio-quiet (RQ) AGN, shows a redder SED, 
broader Balmer lines, stronger [O III] emission, and stronger red wing/weaker blue wing asymmetry of the 
C IV\,$\lambda$1549 emission line. Other authors found that RL and RQ AGN have remarkably similar 
low-ionization emission lines (Mg II and C III]), while high-ionization lines are clearly stronger in RL 
composite spectrum \citep[e.g.,C IV, ][]{francis93,zheng97}. Motivated by these spectral differences in RL 
and RQ AGN, in this work the authors investigate the difference of the BE between the population of RL AGN having the beamed 
continuum emission due to relativistic jet and the sample of RQ AGN.

{The paper is presented as follows. The characteristics of the RL AGN sample and spectroscopic observational 
data are presented in Section,\ref{sec:sample}. The comparison sample of RQ AGN is described in  Section,\ref{sec:RQ}. 
The BE and simulations of the contribution  of non-thermal emission to the BE is presented in Section,\ref{sec:BE}. 
The statistical results and comparison with the RQ samples are presented  in Section,\ref{sec:analysis}, including line-luminosity 
relations, and BE slope differences. The jet contribution to the total non-thermal continuum emission and the 
non-thermal dominance dependence on viewing angle  and equivalent width for Flat Spectrum Radio Quasars (FSRQ) is discussed in Section,\ref{sec:NTD}.  {Finally, discussion and conclusions are presented in  Section,\ref{sec:disycon}}. }

Throughout the paper a flat cosmology model is used with { parameters} $\Omega_{m}=0.3$ ($\Omega_{\Lambda}+\Omega_{m}=1$) and $H_0=70$ km\,s$^{-1}$\,Mpc$^{-1}$.


\section{Sample and Spectroscopic Data of Radio-Loud AGN}
\label{sec:sample}

\subsection{Sample}

The sample of 96 RL AGN studied here is a part of 250 compact extragalactic sources with radio jets (15 GHz) compiled and described by \citet{kovalev05}, that comprises blazars (BL~Lacs and flat-spectrum radio quasars), radio galaxies, and few sources unclassified in the optical regime. AGN of this sample have a core--jet structure on miliarcsecond scales, where the radio jet is aligned close to the line-of-sight. These sources are observed with VLBA at 2~cm \citep[][]{K98,K04,Z02}, and roughly half of the sample is part of the MOJAVE \footnote{\url{http://www.physics.purdue.edu/astro/MOJAVE/index.html}} \citep[``Monitoring of Jets in AGN with VLBA Experiments''; see][]{lister09} program. Most of the sources in the sample have flat radio spectra \citep[$\alpha >-$ 0.5, $F\sim\nu^{+\alpha}$, for $\nu\,>$ 500 MHz;][]{kov99,kov00}, their total flux density at 15\,GHz (obtained in the period 1994--2003) is $>\,$1.5 Jy for Northern hemisphere sources ($\delta >$ 0$^\circ$) and $>2$\,Jy for sources with $-20^\circ<\delta<0^{\circ}$.

Given that 97\% of the sample is comprised by AGN with flat radio spectrum, and broad lines typical of quasars, hereafter the RL AGN sample will be referred to as Flat Spectrum Radio Quasars (FSRQ).

The range in radio-loudness of the FSRQ is 1.2$\leq\,\log\,{R}\,\leq$4.5 with an average value of $\log {R}=\,$3.5. 

{The core--jet structure of the 96 FSRQ makes it a unique sample to study via spectroscopic observations the influence of the jet beaming effects on the broad and narrow  emission line regions (BLR and NLR), and in particular to study the BE in RL AGN.}

\subsection{Spectroscopic data}
\label{subsec:obs}

Optical and ultraviolet spectroscopic data of blazars are presented in full detail in the accompanying spectral 
atlas\footnote{\url{http://vizier.cfa.harvard.edu/viz-bin/VizieR?-source=J/other/RMxAA/48.9}} \citep{torrealba2012}. 
Spectra are available for 123 sources from the MOJAVE/2cm sample \citep[see][]{Torrealba14}, but for the BE analysis 
presented here, the sample was restricted to AGN with $S/N>$10 spectra which involves a sample of 96 FSRQ, which are about half 
of the AGN in the MOJAVE sample. 

As is mentioned in the spectroscopic atlas, the observations 
were acquired at two 2.1~m Mexican telescopes in OAGH\footnote{Observatorio Astrof\'isico Guillermo Haro, in Cananea, 
Sonora, Mexico} and OAN-SPM\footnote{Observatorio Astron\'omico Nacional in San Pedro M\'artir, Baja California,
 Mexico}. In few cases, the spectra were complemented with available databases (HST, SDSS, etc.). Our database 
 is homogeneous in the sense that the same spectral analysis procedures are 
used for fitting the emission lines, de-blending of the Fe II emission and emission-line local continuum fitting. 
To strengthen the analysis results, the flux, line equivalent width, and continuum luminosity 
measurements have not been mixed with data obtained from literature.

Three subsamples of FSRQ were defined:
\begin{itemize}
\item{$\;$The H$\beta$ subsample comprises 18 quasars and 3 radio galaxies. The narrow-line sources with FWHM 
H$\beta\,\lesssim\,1000\,\rm{km}\,s^{-1}$ were excluded. The redshift range is 0.033-0.751 with optical magnitude between 13.6$\,<\,B_{J}\,<\,$18.5.}
\item{$\;$The Mg II\,$\lambda$2798 subsample is the largest data set which comprises 69 quasars. In this case, the redshift 
range is 0.295-2.118 with magnitude between 14.5$\,<\,B_{J}\,<\,$20.6.}
\item{$\;$The C IV\,$\lambda$1549 subsample comprises 31 quasars. The redshift range is 0.295-3.396 with magnitude between 
15.1$\,<\,B_{J}\,<\,$20.9.}
\end{itemize}

It is important to mention that due to the redshift, more than one emission line was available for some sources.

\subsection{Continuum and emission line parameters}
\label{subsec:parameters}

The same spectral analysis procedure was used to measure spectral line parameters (flux and equivalent width) and continuum emission for 
all AGN in our sample. Procedures to obtain the continuum emission and the subtraction of the Fe II contribution are described 
in detail in section~6 of \citet{torrealba2012}. The emission line parameters are measured after subtracting the contribution of Fe II 
emission and a power-law of the local continuum. The spectral range of the data only allows to fit the local continuum with a power-law, 
by selecting regions free of emission or absorption lines. The total emission line flux was measured by Gaussian decomposition of the spectra. 
The decomposition was performed using the task MPFITEXPR from the MPFIT IDL package \citep{mpfit}.

The uncertainty of the emission-line flux is estimated from the formula given in \citet{tresse99} and 
on the average is about 15\,\%. The continuum flux is measured from the iron free spectrum for each AGN 
in the range of $\pm\,$50~\AA. Then the monochromatic continuum luminosities were calculated
$L_{c}\equiv\lambda L_{\lambda}$ at 5100~\AA, 3000~\AA, or 1350~\AA\, 
for the three AGN subsamples.   

The luminosity results of FSRQ samples are:
\begin{itemize}
\item $\;$Ranges of continuum luminosities:  $44.1 \leq \log L_{5100} \leq 46.8$ and $45.6 \leq \log L_{3000} \leq 48$, and $46.3 \leq \log L_{1350} \leq 48.8$.
\item $\;$Mean continuum luminosities: $\log\,L_{5100}=45.7 \pm 0.8$, $\log\,L_{3000}=46.7 \pm 0.5$, and $\log\,L_{1350}=47.6 \pm 0.9$. 
\item $\;$Average uncertainty for $L_c$: 11\,\%, 10\,\%, and 17\,\% for $L_{5100}$, $L_{3000}$, and $L_{1350}$, respectively. 
\item $\;$Mean total line luminosities:  log\,$L_{\rm H\beta}=43.8 \pm 0.8$, log\,$L_{\rm Mg\,II}=44.8 \pm 0.5$, and log\,$L_{\rm C\,IV}=45.8 \pm 0.7$.
\end{itemize}

The equivalent width for each emission line was calculated using the ratio of the total line luminosity ($L_{line}$) and monochromatic continuum multiplying by the wavelength associated with the corresponding continuum, $W_\lambda= (L_{line}/L_{c})\times\lambda$. The $W_\lambda$ uncertainties are about $30-35$\,\% depending on the mean signal-to-noise ratio of the spectrum.

Uncertainties of the equivalent widths $W_\lambda$ for the emission lines near 5100\,\AA\ with mean spectral ${\rm S/N}\sim 15$ lie in the range 10$\,-\,$15\%. Near the 3000 \AA\, region, the uncertainties are roughly 12\,\% with  S/N$\,\sim\,$20, and for  $W_\lambda$(C IV) the average uncertainty is $\sim$14\,\% with S/N$\,\sim\,$15. 

The mean $W_\lambda$ and its standard deviation of H$\beta$, Mg II, and C IV 
emission lines are $(76.6 \pm 23.8)$~\AA, $(42.4 \pm 21.8)$~\AA, and $(27.0 \pm 14.4)$~\AA, respectively.


\section{Samples of Radio-Quiet AGN}\label{sec:RQ}

To compare the BE in our FSRQ sample, two samples of RQ AGN were selected. For H$\beta$ the sample used comes from \citet{greene05} 
while for Mg II and C IV the sample comes from \citet{shen11}. This control sample was compared to the BE in FSRQ. 
Both samples of RQ AGN were selected from the Sloan Digital Sky Survey \citep[SDSS, ][]{york00}.

The RQ control sample for H$\beta$ emission consists of 229 RQ AGN from the Third Data Release \citep[DR3, ][]{aba05} 
with $z \leq 0.35$. The second sample is taken from The Seventh Data Release  \citep[DR7, ][]{aba09} with  
44,000 quasars having the Mg II emission line (0.35$\leq z \leq$2.25), and 10,000 quasars with C IV 
emission line (1.5$\leq z \leq$4.95). Both samples are assumed to be dominated by a population of RQ AGN 
\citep[e.g., ][]{shaw12}.  

\citet{greene05} and \citet{shen11} use the following procedure to measure the spectral line and continuum characteristics. 
They decompose the spectrum for each source by simultaneous fitting of two-component model 
consisting of featureless continuum and the empirical Fe II template from \citet[][H$\beta$ region]{boroson92} 
and \citet[][Mg II region]{VW01}. \citet{shen11} fitted the local continuum with a single power-law 
in the wavelength intervals between 2200-2700 \AA\, and 2900-3090 \AA\, near the Mg II line and 1445-1465 \AA\, and 
1700-1705 \AA\, near the C IV emission line. The featureless continuum in the region of H$\beta$ line 
was approximated by a  double power-law broken at 5000 \AA\, under the requirement that the combined flux 
of the two components at $\sim$5600 \AA\, (near H$\beta$) be equal to the observed flux at that point \citep{greene05}. 
\citet{shen11} measured the C IV line flux without iron subtraction which may lead to an overestimation 
of $W_{\lambda}$ by $\sim$0.05 dex on average. In the RQ samples, the emission line profile is modeled 
as a multicomponent Gaussian taking into account both the broad and narrow components.


\section{Baldwin Effect}\label{sec:BE}

\subsection{$W_{\lambda}$ vs. $L_{c}$}

The relation between the emission line equivalent width ($W_\lambda$) and the continuum emission
 luminosity ($L_{c}$) is given by \citet{baldwin77},
\begin{equation}
  \log\,W_\lambda = \alpha\,+\beta\,\log\,L_{c}.
  \label{ec:baldwin}
\end{equation}
The slope $\beta$ is found to be negative for RQ AGN \citep{baldwin77,shields07}, i.e., 
the equivalent width of the emission line (or the contrast between the line and continuum luminosities) 
decreases towards large continuum luminosities. 

Equation~(\ref{ec:baldwin}) can be transformed to a relation between the total line luminosity $L_{\rm line}$ 
and the monochromatic continuum luminosity $\lambda L_{\lambda} \equiv L_{c}$ measured at a certain
wavelength $\lambda$,

\begin{equation}
 \log\,L_{\rm line}=A\,+B\,\log\,L_{c},
\label{ec:line}
\end{equation}
by replacing $\alpha$ and $\beta$ with,
\begin{equation}
 \alpha =A + \log\,\lambda
\label{ec:alfa}
\end{equation}
\begin{equation}
 \beta = B - 1,
\label{ec:beta}
\end{equation}
and considering that $W_\lambda\simeq \,L_{\rm line}/L_{\lambda}$.

\subsection{Contribution of non-thermal emission to the Baldwin Effect}
\label{ssec:non-thermal-BE}

The optical continuum emission in RQ AGN is assumed to be isotropic and generated in the accretion disk, 
so that the continuum luminosity is $L_{c} = L^{\rm RQ}_{\rm disk}$. On the other hand, for FSRQ the 
optical continuum emission has two components,
the thermal emission from the disk ($L^{\rm BL}_{\rm disk}$) and the beamed non-thermal emission from the 
relativistic jet ($L_{jet}$), i.e. 
$L_{c} = L^{\rm BL}_{\rm disk} + L_{\rm jet}$. It was assumed that the main contribution to the broad line emission 
is attributed to the disk thermal emission, while the beamed emission from the jet is produced beyond the BLR and, hence, has no 
contribution to the Broad Line (BL) emission. 

\begin{figure*}[htbp]
\linespread{1.25}
\begin{center}
\includegraphics[width=0.49\linewidth]{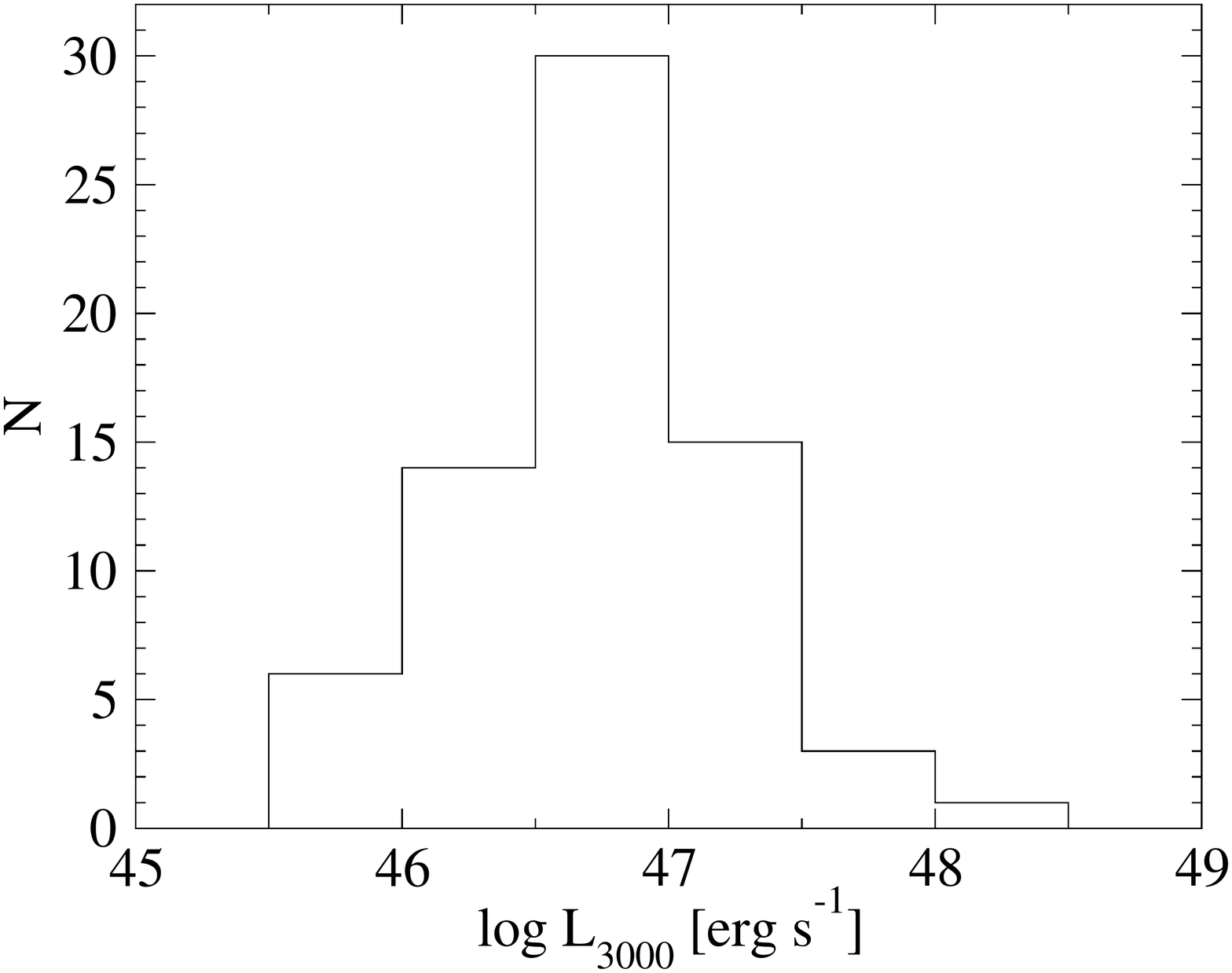}
\includegraphics[width=0.49\linewidth]{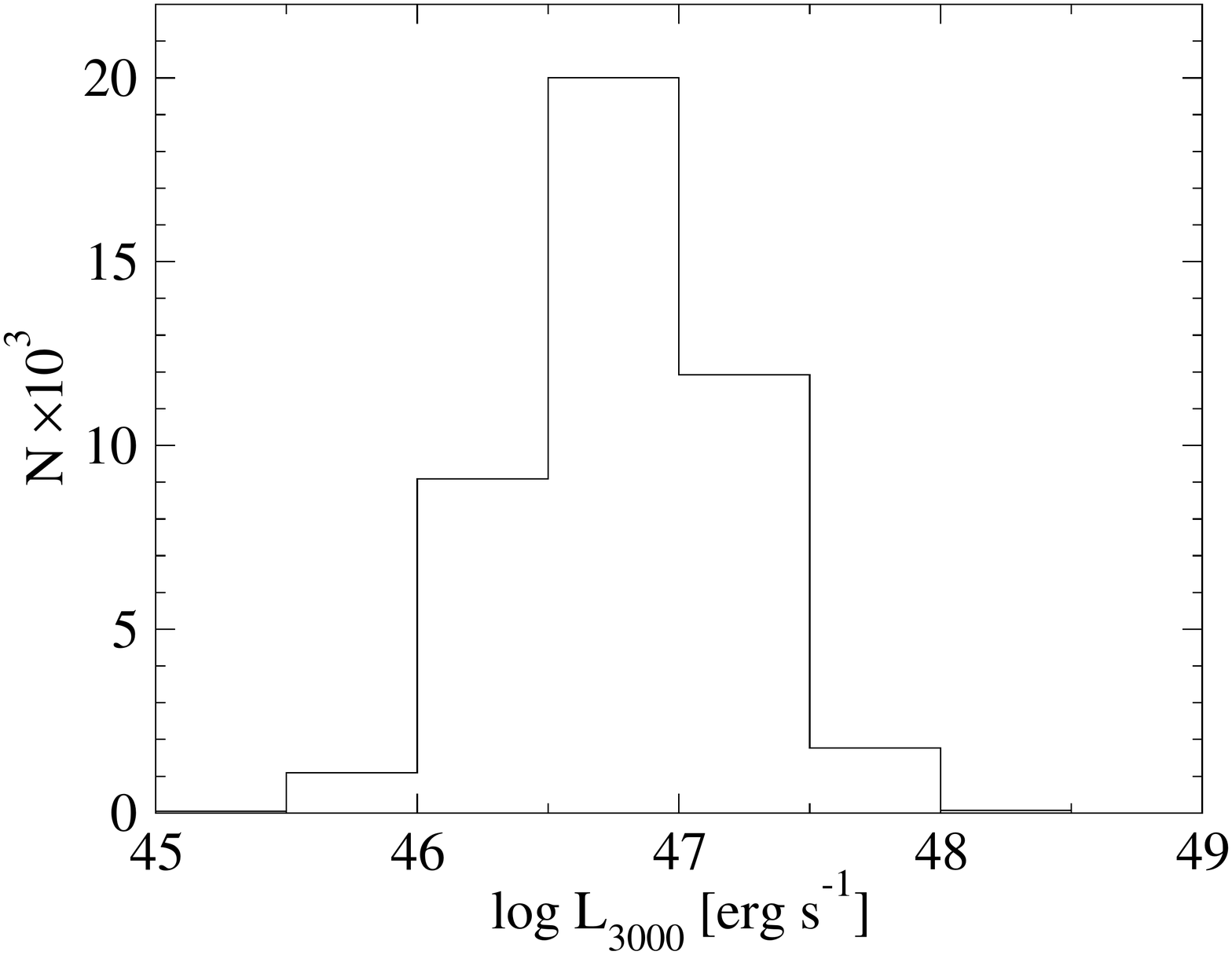}
\caption{Distributions of continuum luminosity at 3000\AA$\;$ obtained for the observed FSRQ sample (left panel); and for the simulated RQ sample (right panel).}
\label{fig_conthist}
\end{center}
\end{figure*}

\begin{figure*}[h]
\linespread{1.25}
\begin{center}
\includegraphics[width=\linewidth]{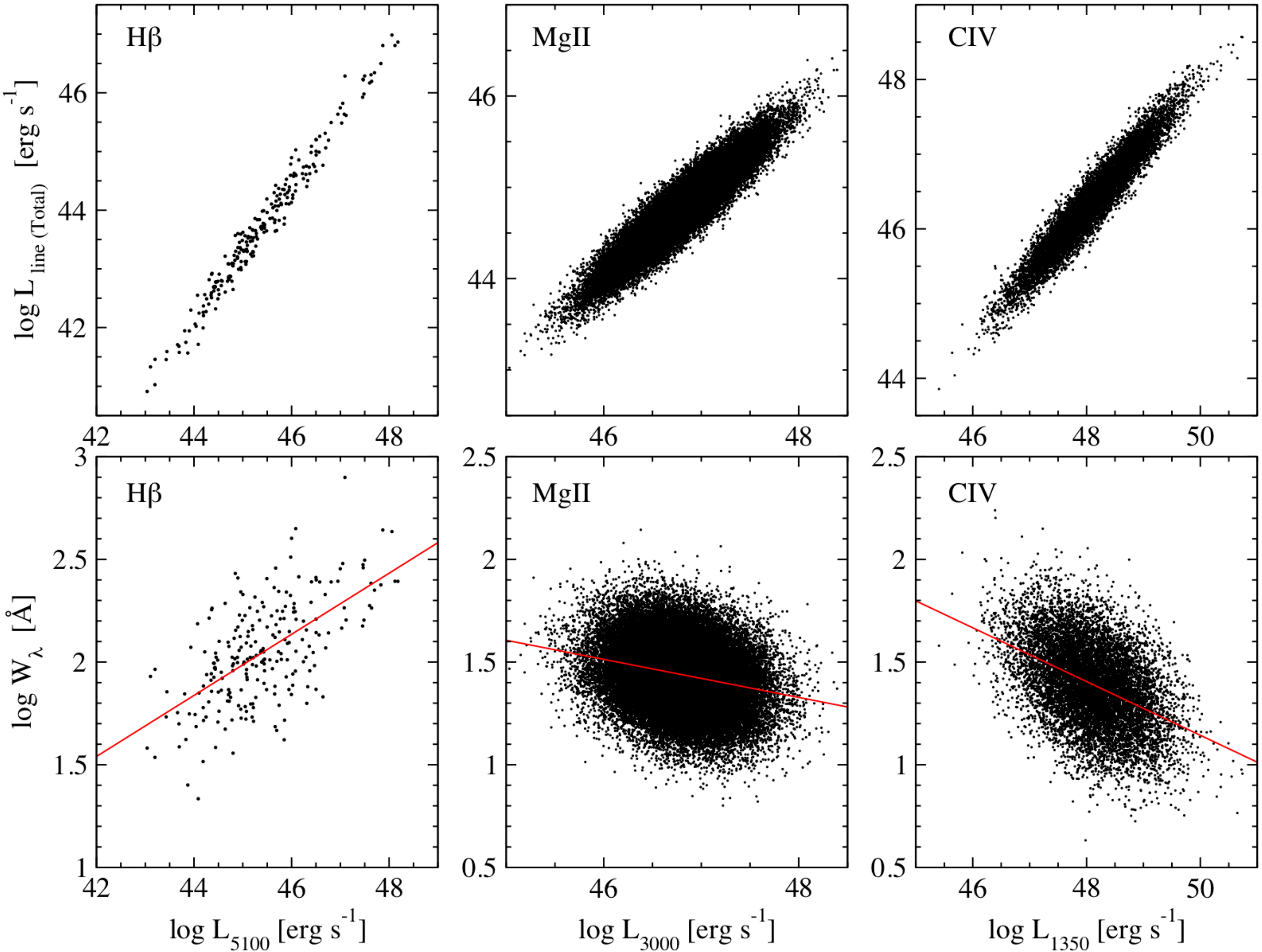}
\caption{Relations obtained for the H$\beta$ (left panel), Mg II (middle panel), and CIV (right panel) emission lines. Top row: $L_{line}$-$L_c$ plot with simulated data. Bottom row: $EW$-$L_c$ plot with simulated data.}
\label{fig_simurq}
\end{center}
\end{figure*}

A simulation was performed in order to estimate the statistical properties of the RQ samples studied by \citet{greene05} and \citet{shen11}. 
The purpose is to compare them to the RL sample presented in this paper.
First. a distribution of continuum luminosity was taken (see Figure~\ref{fig_conthist}), based on the luminosity distributions observed for our sample of FSRQ. 
For all three lines a gaussian distribution represents well the data. Then, using the line luminosity - continuum luminosity relations 
described in the afore mentioned papers, and the scatter obtained for these relationships, representative line luminosities 
of the sample of RQ AGN were generated. Using these line luminosities and the assumed distribution of continuum luminosities, 
the equivalent widths for the simulated RQ sample were calculated. The number of simulated values in each case is equal to the number 
of data points in the original RQ samples. Figure~\ref{fig_simurq} shows the simulation results for the H$\beta$, Mg II, and CIV lines.

After generating the data, it was separated in order to match the continuum luminosity range on our sample of FSRQ. 
Using the simulated data that fall in our observed range, the mean 
and standard deviation of the equivalent width for the three lines H$\beta$, C IV and Mg II, were calculated. It is 
worth mentioning that for C IV  
and Mg II, all the simulated data fall inside the desired continuum luminosity ranges.  

Then, a linear least-squares algorithm in one dimension was applied, using the IDL task 
FITEXY\footnote{\url{http://user.astro.columbia.edu/~williams/mpfitexy/}}, to perform a linear fitting to the 
$L_c$\,-\,$W_{\lambda}$ relation to obtain the Baldwin Effect of the RQ sample. 

From these simulations, the next conclusions were drawn:
\begin{itemize}
	\item{$\;$The slope and uncertainty on the simulated L$_c$\,-\,$W_{\lambda}$ relation remains unchanged, regardless of the input continuum luminosity distribution used.}
	\item{$\;$The changes on the intercept and uncertainty on the simulated L$_c$\,-\,$W_{\lambda}$ relation, with changes on the input continuum luminosity distribution are negligible.}
	\item{$\;$The mean and standard deviation of the simulated equivalent widths can change drastically, depending on the continuum luminosity distribution used as input.}
\end{itemize}

%


\section{Statistical Analysis}
\label{sec:analysis}

To understand the effect of a non-thermal emission in the BE for blazars, it is necessary to analyze the difference in the 
$L_{\rm line}-L_{\rm c}$ and $W_{\lambda}-L_{\rm c}$ relations between the samples of FSRQ and RQ AGN. 

\begin{figure*}[htbp]
\linespread{0.5}
\centering
   \includegraphics[width=1\linewidth]{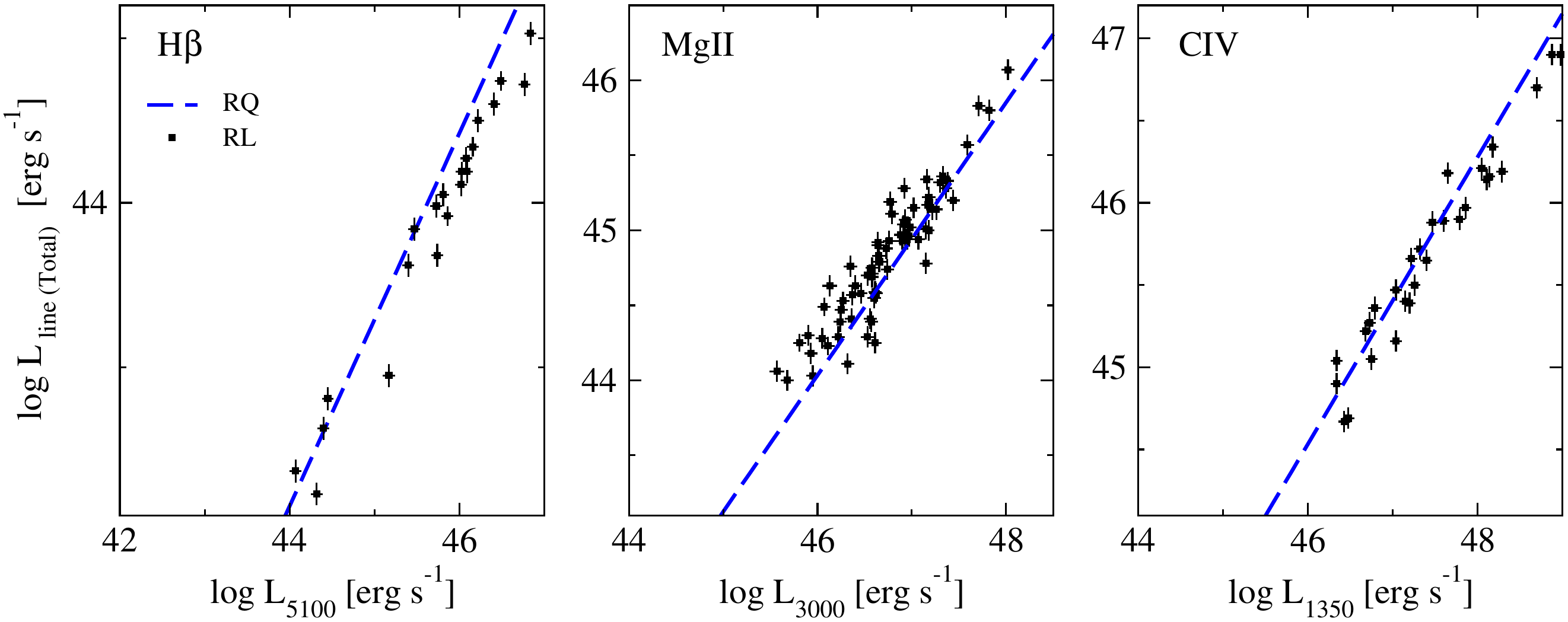}
\caption{\label{fig_lc_ll}{Emission line luminosity against continuum luminosity of FSRQ (squares): $L_{\rm H\beta}$ vs. $L_{5100}$ (left panel), $L_{\rm Mg\,II}$ vs.  $L_{3000}$ (middle panel), and $L_{\rm C\,IV}$ vs.  $L_{1350}$ (right panel). The dashed line reproduces the relations $L_{{\rm H}\beta}-L_{5100}$ (left panel) from \citet{greene05}, $L_{\rm Mg\,II}-L_{3000}$ and $L_{\rm C\,IV}-L_{1350}$ (middle and right panels) from \citet{shen11}. }}
\end{figure*}

\subsection{Comparison of line-luminosity relations}
\label{subsec:comparisonLL}

The relation between the line luminosity and the continuum luminosity of RQ AGN are derived for H$\beta$ by
 \citet{greene05}, and  for Mg II and C IV by \citet{shen11}, using the weighted linear fitting of binned data for
the {total flux} of the emission lines H$\beta$, Mg II, and C IV, and their respective continuum luminosities at
5100~\AA\, 3000~\AA\,, and 1350~\AA\, (dashed lines in Figure~\ref{fig_lc_ll}). The slope and {intercept} of their fittings are presented in the top part of Table~\ref{tab:line_lum}. 

The emission line and the corresponding continuum luminosity data of FSRQ are shown for H$\beta$, Mg II, and C IV emission
lines in Figure~\ref{fig_lc_ll}. The same fitting procedure as in \citet{greene05} and  \citet{shen11} were followed, and the relation defined 
in \S~\ref{subsec:obs} between the line luminosity and the continuum luminosity for each subsample 
of FSRQ (straight lines in top panels of Figure~\ref{fig_line_c_EW}) was derived. Fitting parameters of our subsamples and significance of correlations
between line and continuum luminosities are presented in the lower part of Table~\ref{tab:line_lum}. 

Significant correlations for all line luminosities at the confidence level of $\ge 98.8\,\%$ were found. 
It is noticeable that the slopes $B$ of line-continuum luminosity relations measured for FSRQ have a tendency to be shallower for the three ions 
H$\beta$, Mg II, and C IV, than those found for RQ AGN, see Figure~\ref{fig_line_c_EW} (top panels) and Table~\ref{tab:line_c_EW}. 
In order to quantify the significance of the difference in the slopes of the $L_c$-$L_{line}$ relations for RQ and FSRQ; 
an unpaired $t$-test\footnote{\url{http://graphpad.com/quickcalcs/ttest1/}} was applied.
For the H$\beta$ line, the two-tailed P value is 0.0057, corresponding to a statistically significant difference. 
For the Mg II line, the two-tailed P value is 0.0263, corresponding also to a statistically significant difference. 
However, for the C IV line, the two-tailed P value is 0.4370, corresponding to a non-statistically significant difference. 
The authors suggest that the differences found for Mg II and H$\beta$ are indicating the contribution of an extra emission line component, possibly related to the jet.

\subsection{Baldwin effect comparison}
\label{subsec:comparisonBE}

\begin{figure*}[htbp]
\linespread{1.0}
\centering
\includegraphics[width=1\linewidth]{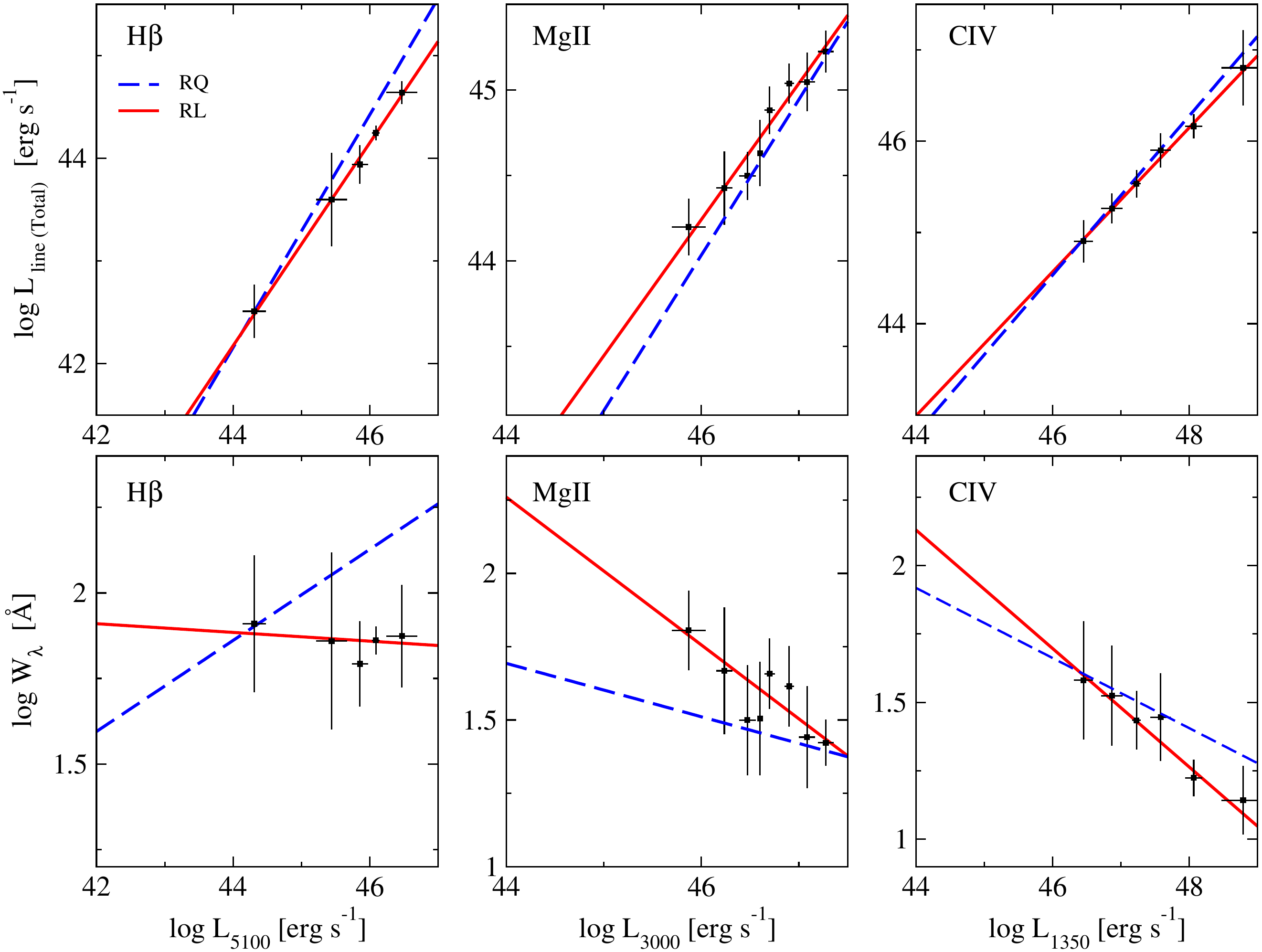}
\caption{\label{fig_line_c_EW}{The top panels shows the continuum luminosity associated to each line ($L_{c}$) and the total line luminosity $L_{line}$ for our binned data. The bottom panels shows $L_{c}$ and the equivalent widths ($W_{\lambda}$) estimated from the data in the former panels. The solid line is the best weighted linear fit to our data. The dashed line are derived from the simulations described in \S~\ref{ssec:non-thermal-BE} using  the relations $L_{{\rm H}\beta}-L_{5100}$  from \citet{greene05}, $L_{\rm Mg\,II}-L_{3000}$ and 
$L_{\rm C\,IV}-L_{1350}$  from \citet{shen11} to derive the corresponding line luminosities and equivalent widths.}} 
\label{figLL-LEW}
\end{figure*}

The Baldwin Effect of FSRQ is derived by a weighted linear fitting to the binned data, taking into account uncertainties in both axes using the IDL task FITEXY. 
Bins of the data were set along the $L_{\rm c}$ and measure the mean and standard deviation of $W_\lambda$ in each bin. 
It should be noted that an adaptive data bin was used, in order to get the same number of measurements in each bin. 
The weighted fit lines for each H$\beta$, Mg II, and C IV lines are presented in Figure~\ref{fig_line_c_EW} (full lines in the lower panels) 
and their fitted parameters are listed at the bottom part of Table~\ref{tab:line_c_EW}.

For each emission line, simulated values of $W_\lambda$ and $L_{\rm c}$ for RQ AGN (see Figure~\ref{fig_simurq}) are used to generate the data set, 
which is then fitted by the weighted linear method described above (dashed lines in bottom panels of Figure~\ref{fig_line_c_EW}). The slope, intercept, 
and uncertainties for each line, are presented in the top part of Table~\ref{tab:line_c_EW}.

There is a difference in the slopes for the BE for the simulated RQ sample and the observed FSRQ sample; 
as can be seen in the bottom panels of Figure~\ref{figLL-LEW}. The significance of these differences is tested by means of an unpaired $t$ test. 
For H$\beta$ the test results in a statistically significant difference, with a P-value of 0.0007. 
For Mg II, the test also results in a statistically significant difference, with a P-value of 0.0016. 
However, for C IV, the test results in a non-statistically significant difference, with a P-value of 0.1161.

\begin{figure*}[htbp]
\linespread{0.1}
\centering
\subfigure[]{\includegraphics[width=\textwidth]{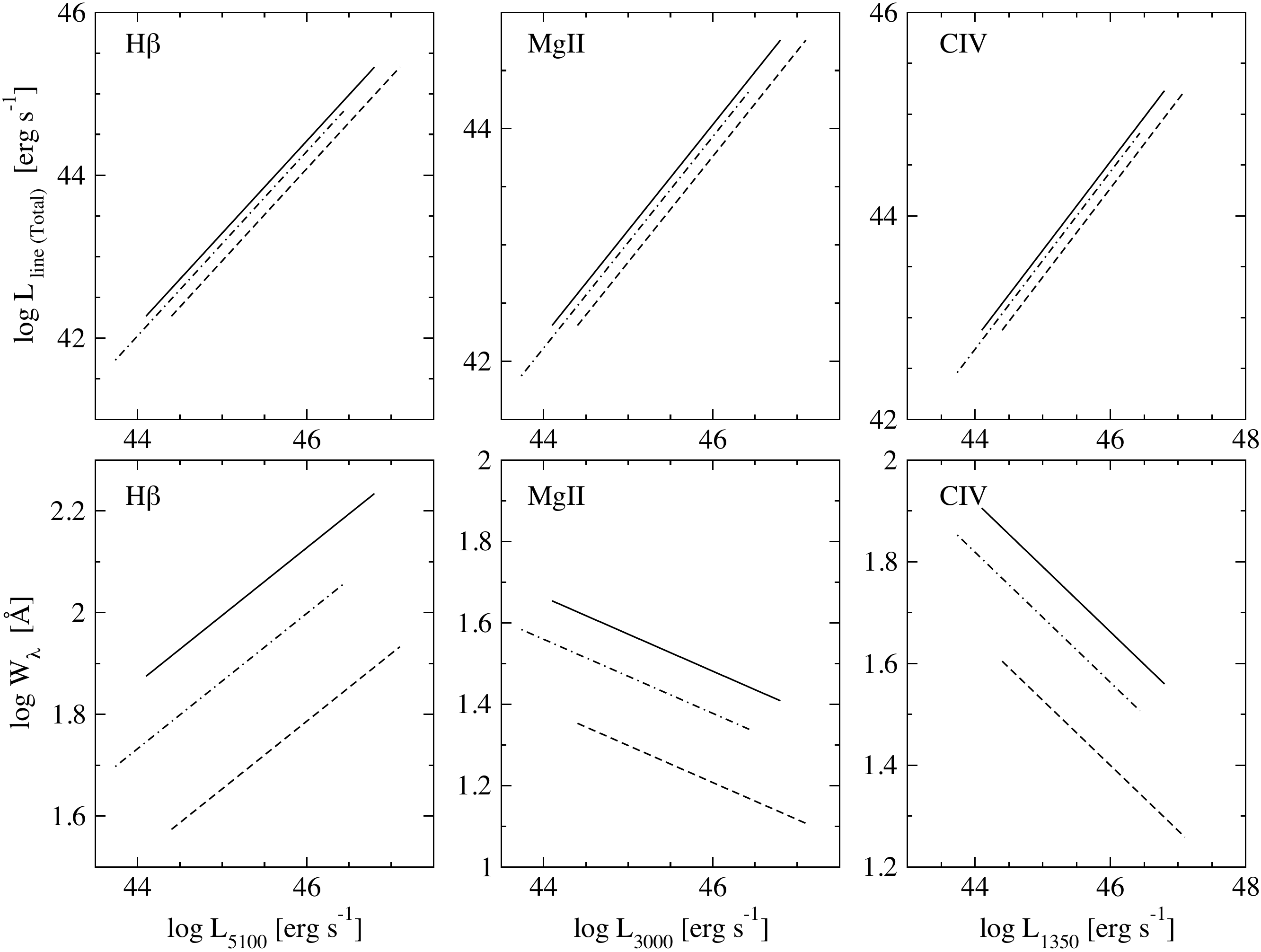}}   \\
\caption{Simulations of $L_{line}$-$L_c$ relation (top panels) and  $W_{\lambda}$-$L_c$ relation (bottom panels) for the H$\beta$ (left columns), 
Mg II (middle columns), and C IV (right columns) emission lines for three different scenarios: RQ (solid line), with radio jet component (dashed line), 
jet and disk components decreased, each by a different factor (dot-dashed line). }
\label{fig_sim2}
\end{figure*}

In order to explain the differences between the slopes in both the relations $L_{cont}-L_{line}$ and $L_{cont}-W_{\lambda}$ of the RQ \citep{greene05,shen11} 
and the FSRQ (our sample), a simulation was designed showing the behavior of the Baldwin Effect under the accepted paradigm of RQ and RL AGN. The simulation results for three different scenarios are listed:
\begin{itemize}
\item $\;$First a RQ system was simulated, using as base the RQ relations for the three lines published by the above authors (solid line in Figure~\ref{fig_sim2}).
\item $\;$Then a continuum component (simulating the jet) was added of the same luminosity as the disk component. The emission line is 
calculated using only the continuum component from the accretion disk (dashed line in Figure~\ref{fig_sim2}).
\item $\;$Then both continuum components were decreased, each by a different a factor. This with the aim of probing if the differences in variation amplitudes of the disk and the jet, are responsible for the differences observed in the slopes. Many different variability factors were simulated, however, all of them just result in a Y-axis shift of the model.
Again, the emission line is calculated using only the continuum component from the accretion disk (dot-dashed line in Figure~\ref{fig_sim2}).
\end{itemize}

The simulation results for the H$\beta$, Mg II, and C IV lines are shown in Figure~\ref{fig_sim2}.

As evidenced by the simulations presented above, the presence of an additional continuum component produced by the jet, does not explain the BE slope difference found between FSRQ and RQ AGN; the change produced is only a parallel Y-axis shift and not a change in slope.


\section{Non-thermal Contribution to the Continuum Emission}
\label{sec:NTD}

\subsection{Non-thermal dominance for FSRQ}

To quantify the contribution of the jet emission to the total optical/UV emission, the non-thermal dominance (\textit{NTD}) 
introduced in \citet{shaw12}, was estimated. They defined the $NTD$ as
\begin{equation}
NTD\,=\,\frac{L_{obs}}{L_{p}},
\label{ec:ntd}
\end{equation}
where $L_{obs}$ is the observed continuum luminosity and $L_{p}$ is the predicted continuum luminosity estimated from the emission-line luminosity for a non-blazar sample. It is assumed that the broad-line emission reflects the thermal power of the accretion disk. 

In this work, the authors define an alternative $NTD$ for FSRQ: 
\begin{equation}
NTD\,=\,\frac{L_{obs}}{L_{p}} \,=\, \frac{L_{disk} + L_{jet}}{L_{p}},
\label{ec:ntd2}
\end{equation}
where $L_{obs}$ is the observed continuum luminosity, $L_{p}$ is the predicted disk continuum luminosity estimated from the emission line, $L_{disk}$ is the continuum luminosity emitted by the accretion disk, and $L_{jet}$ is the jet contribution to the continuum luminosity. If the emission line is only affected by the disk $L_{p}\,=\,L_{disk}$, so that
\begin{equation}
NTD\,=\,1 + \frac{L_{jet}}{L_{disk}},
\label{ec:ntdj}
\end{equation}
which shows that $NTD \geq 1$. Note that $NTD=1$ means that the continuum is due only to thermal emission, $NTD>1$ shows that a superluminal jet exists that contributes to the continuum luminosity, and $NTD>2$ means that ${L_{jet}}>{L_{disk}}$.

The emission line luminosity vs. the observed continuum luminosity for the three subsamples is presented in the top panels of Figure~\ref{fig_line_c_EW}. It was found that the majority of blazars from the H$\beta$ (81\%) and C IV (84\%) subsamples are located below the fiducial relations found by \citet{greene05} and \citet{shen11} presented as dashed lines. Thus, these FSRQ must have a significant non-thermal contribution in the optical and UV bands, i.e. $NTD>1$ for most sources. The excess emission is likely to be a boosted optical/UV emission from the relativistic jets, and it was expected that all quasars in our subsamples should have $NTD>1$. 

\begin{figure*}[htbp]
\linespread{0.5}
\centering
\includegraphics[width=0.9\textwidth]{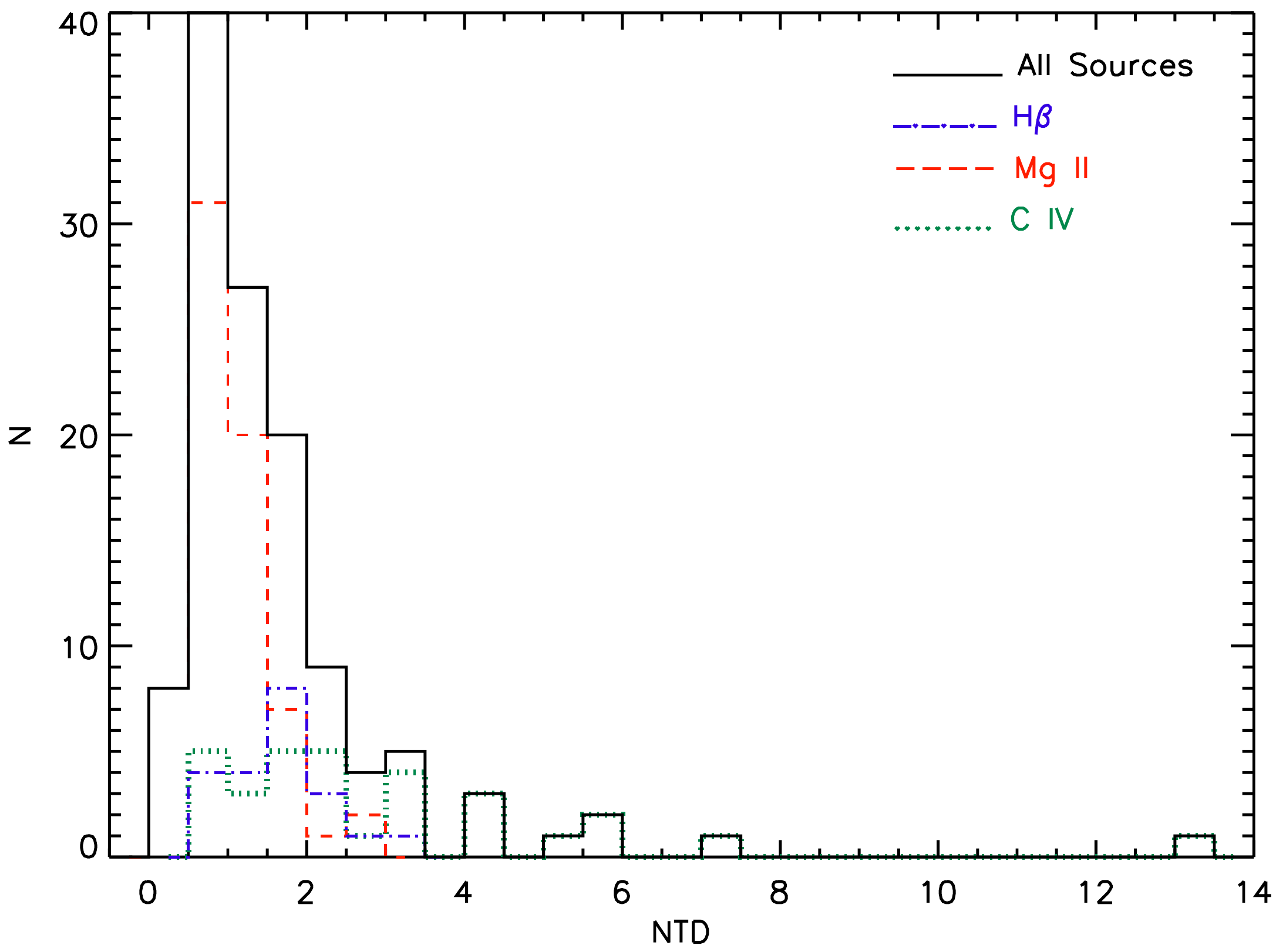}   \\
\caption{Histogram of the values obtained for the non-thermal dominance ($NTD$). The different emission lines are indicated in different colors and line styles. H$\beta$ in blue (dot-dashed), Mg II in red (dashed), C IV in green (dotted). The black solid line represents the sum of the three emission lines.}
\label{fig_ntd_hist}
\end{figure*}

Figure~\ref{fig_ntd_hist} shows the distribution of $NTD$ values obtained for the three emission lines of our study; 
it is noticeable that the peak is around $NTD=0.5-1.0$. Further results concerning the $NTD$ are discussed in the next section. 
Instead of the expectation, the $NTD$ obtained for our sample of FSRQ shown in the vertical axis of Figure~\ref{fig_NTD_VA} 
spans values in the regions $NTD>2$, where ${L_{jet}}>{L_{disk}}$ in Equation~(\ref{ec:ntdj}). Few sources have $NTD>2$. 
On the other hand the region $NTD>1$ is populated by a large number of sources, where the superluminal jet contributes to the continuum luminosity, 
and a number of sources have $NTD<1$. 

{ 
It was found that 56\% of the Mg II subsample (filled dots in Figure~\ref{fig_NTD_VA}) have $NTD<1$, while only 
19\% of H$\beta$ (empty squares in Figure~\ref{fig_NTD_VA}) and 16\% of C IV sources (filled triangles in Figure~\ref{fig_NTD_VA}) 
have $NTD<1$. {Values of $NTD<1$, were found as well}  by \citet{shaw12}. {This result probably means that, not only an additional component 
of the BLR (BLR2) exists together with the canonical BLR, which can be related to or activated by the jet (c.f. $L_{p(BLR2)}$ in Equation \ref{ec:ntdjblr})}; 
but also that it scales differently with the continuum luminosity than the canonical BLR component ($L_{p(BLR2)}>L_{jet}$). The reason 
why this affects more the Mg II subsample is unclear, but could be due to a larger sample than the other two 
ion subsamples, or a possible ion stratification in the BLR where Mg II zones are closer to the ionization 
source in the inner part of the jet. Observational evidence for the presence of BLR material 
located at parsec scales down to the radio core has been found by coordinated spectroscopic and VLBI 
monitoring studies.  More specifically, \citet{arshakian10_b} and \citet{tavares10} found evidence for BLR material 
around the radio core for the radiogalaxies 3C 390.3 and 3C 120, respectively.
}

In order to explore this possibility, it was assumed that there exists an emission line component related to the jet then $L_{p}>L_{disk}$ therefore:
\begin{equation}
NTD\,\neq\,1 + \frac{L_{jet}}{L_{disk}}.
\label{ec:ntdjj}
\end{equation}
If the predicted continuum luminosity obtained from the emission line component related to the disk is called $L_{p(BLR1)}$, and the predicted continuum luminosity obtained from the emission line component related to the jet is called $L_{p(BLR2)}$, then:
\begin{equation}
NTD\,=\, \frac{L_{disk} + L_{jet}}{L_{p}}\,=\,\frac{L_{disk} + L_{jet}}{L_{p(BLR1)}+L_{p(BLR2)}},
\label{ec:ntdjblr}
\end{equation}
where $L_{p(BLR1)}=L_{disk}$. If $L_{p(BLR2)}>L_{jet}$  then it is possible to obtain values of $NTD<1$.

The finding of values $NTD<1$, most specially in Mg II (56\% of FSRQ), seems to support the idea of a BLR component related to the jet. This analysis also suggests that the emission line component related to the jet scales differently with the continuum, than the canonical broad line region scaling.

\citet{arshakian10} and \citet{torrealba11} showed that, for the MOJAVE blazars, optical (5100\,\AA) and radio VLBA total emission at 
15\,GHz ($L_{\rm VLBA}$) are correlated on milliarcsecond scales. They suggest a synchrotron origin of radio and optical emission for quasars and BL 
Lacs which is boosted by the relativistic jet. 
 
Application of partial Kendall's $\tau_{p}$ statistical analysis\footnote{Partial Kendall's $\tau_{p}$ rank 
correlation removes the common dependence of luminosities on redshift.} to the Mg II subsample shows that 
$L_{3000}$ and $L_{\rm VLBA}$ are correlated at a $c.l.=$ 99.9\%  ($\tau_{p}=$ 0.27). While, for the C IV subsample, the correlation 
between  $L_{1350}$ and $L_{\rm VLBA}$ is not significant  ($\tau_{p}=$ 0.21and $c.l.=$ 91.4\%), but the correlation recovers for 
the relation between $L_{1350}$ and jet luminosity with $\tau_{p}=$ 0.20 and $c.l.=$ 97.3\%. Note that  $L_{jet}$ is equal to the difference 
between total VLBA and radio core luminosities \citep[see][]{arshakian10}. These correlations indicate also that the bulk of the UV emission 
is non-thermal and, most likely, produced in the jet, which also supports the contribution of a jet-BLR component to the continuum luminosity.
 
Other evidence for a non-thermal origin of the variable optical emission comes from the link between the jet 
kinematics on sub-parsec scales and optical continuum flares on scales from few months to few years
\citep{Perez89,arshakian10_b,tavares10,tavares13}. 
These findings suggested that the source of the non-thermal variable optical emission is located in the innermost part 
of the sub-parsec scale jet, {which is a region that may be close to the BLR clouds and thus possibly affecting and activating it.  }

\subsection{Dependence of non-thermal contribution on jet viewing angle}
\label{NTDtheta}

The viewing angle of the jet ($\rm \theta_{var}$) was estimated using the variability Doppler factor ($\rm \delta_{var}$) and apparent speed 
of the jet  $\rm \beta_{a}$ \citep[in units of the speed of the light; e.g.,][]{LV99}:

\begin{equation}
\label{ec:angulo_jet}
\rm \theta_{var}=\rm arctan \dfrac{2\beta_{a}}{\beta_{a}^{2}+\delta_{var}^{2}-1}.
\end{equation}

Recent values of $\rm \beta_{a}$ are taken from the MOJAVE 
website\footnote{\url{http://www.physics.purdue.edu/astro/MOJAVE/index.html}} and $\rm \delta_{var}$ from \citet{H09}.
The latter parameter is available for 35 sources from the sample of 96 AGN. For the remaining objects, the empirical relation between the jet 
viewing angle ($\rm \theta_{j}$) and  the total radio luminosity at 15\,GHz was used
($\rm L_{VLBA}$) obtained for 62 blazars from the statistically complete MOJAVE-1 sample, c.f. equation (11) in  \citet{arshakian10}:

\begin{equation}
 {\log(\theta_{j})\,=\,(7.92\pm0.78)\,+\,(0.26\pm0.03) \log L_{\rm VLBA}}
\label{eq:theta}
\end{equation}

Note that the range of $L_{\rm VLBA}$ of our sample ($42.1\leq\,\rm \log L_{VLBA}\,\leq\,46.0$) is similar to the one in \citet{arshakian10}.

Viewing angles $\rm \theta_{var}$ of 71 blazars from the MOJAVE/2cm were estimated by \citet{H09} using the variability Doppler factors and apparent 
speeds of the jets. Note that errors of viewing angles cannot be estimated because of difficulties and significant uncertainties in Doppler factor values ($\rm \delta_{var}$) associated with each source \citep{arshakian10}. 

Seventy one values of $\rm \theta_{j}$ from Equation~(\ref{eq:theta}) were computed and compared with independent measurements of $\rm \theta_{var}$ 
in Equation~(\ref{ec:angulo_jet}). The Spearman rank correlation between the two samples is $\rho=0.56$ with $c.l.>99.99\%$ 
($P=$3$\times$10$^{-7}$), indicating that the measurements 
of $\theta_{\rm j}$ are statistically reliable for viewing angles larger than $\sim$1 degree.

\begin{figure}[htbp]
\linespread{1.0}
\centering
  \includegraphics[width=1\linewidth]{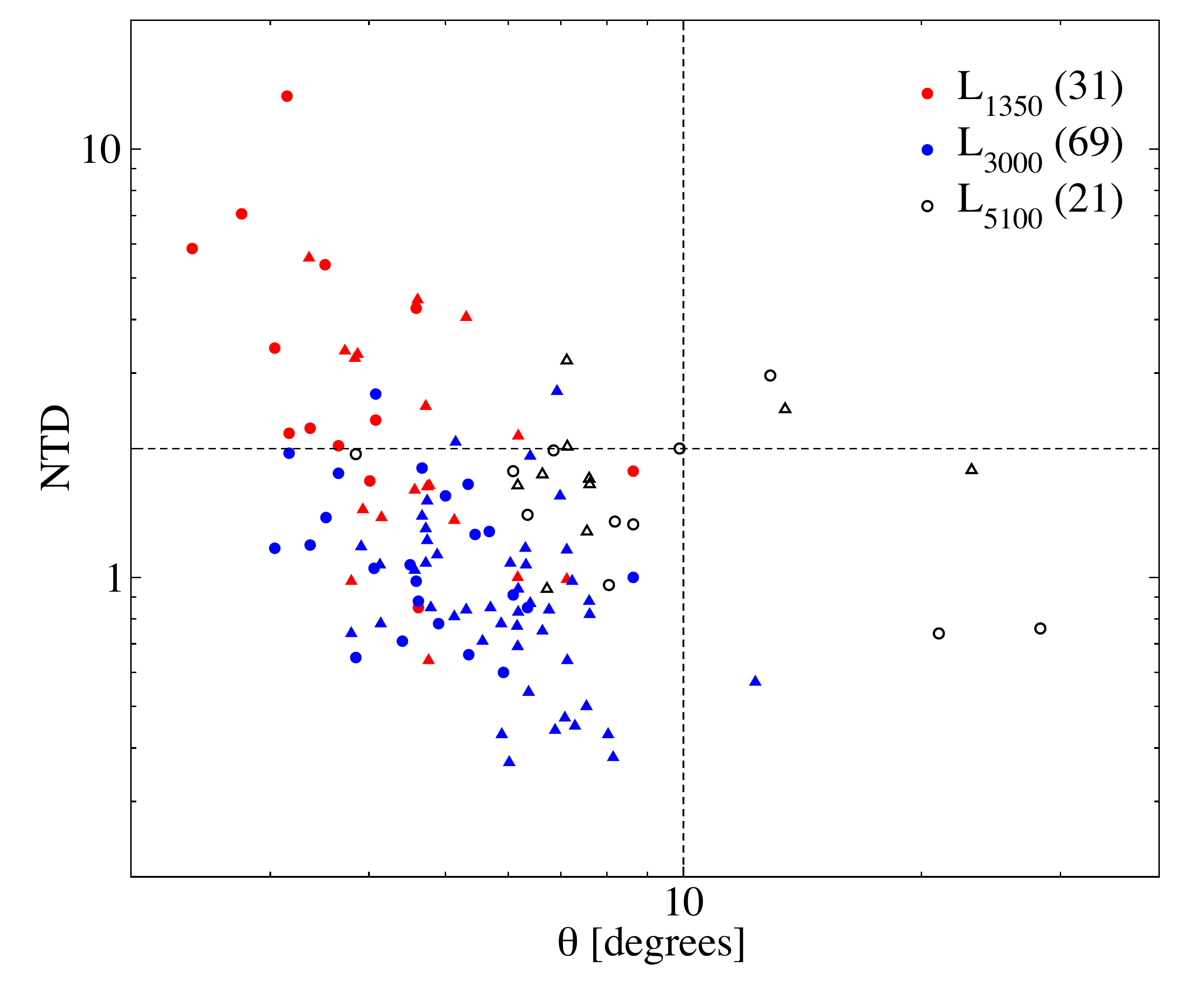}
\caption{Non-thermal dominance (ratio between the observed and the 
predicted continuum emission) vs. viewing angle of the jet.  The colors refer to blazars observed at 1350\,\AA\, (red; C IV), 3000\,\AA\, 
(blue; Mg II), and 5100\,\AA\, (empty; H$\beta$). The circles refers to blazars with $\theta$ calculated using measurements of $\beta_{app}$ and $\delta_{var}$, while the triangles are blazars with $\theta$ calculated through Equation~\ref{eq:theta}. Vertical and horizontal  dashed lines mark the jet viewing angle 
of $\theta$=10$^{\,\circ}$ and $NTD$=2, respectively.}
\label{fig_NTD_VA}
\end{figure}

The viewing angles $\theta_{\rm j}$ of 96 FSRQ were used to analyze the $NTD-\theta$ relation plane presented in Figure~\ref{fig_NTD_VA}. 
There is a negative trend between $NTD$ and viewing angle of the jet for the majority of blazars with viewing angles less than $10^{\circ}$. 
For these sources, the Kendall's partial correlation is $\tau_p=-0.05$ with probability $P=7.6\,\times\,10^{-5}$ ($c.l.$ of 99.99\,\%) indicating for a 
significant negative correlation between jet viewing angle and $NTD$.
This correlation is mainly due to quasars of the Mg II subsample. No significant correlation is found for the H$\beta$ and C IV subsamples, most likely, because of their smaller sampling.

In this work, the authors conclude that the Mg II subsample shows that the non-thermal dominance $NTD$ of the optical and UV continuum emission decreases with viewing angles of the jet, in agreement with the prediction of the relativistic beaming theory.

\subsection{Dependence of equivalent width on jet viewing angle}
\label{NTDtheta}

\begin{figure*}[htbp]
\linespread{1.0}
\centering
  \includegraphics[width=1\linewidth]{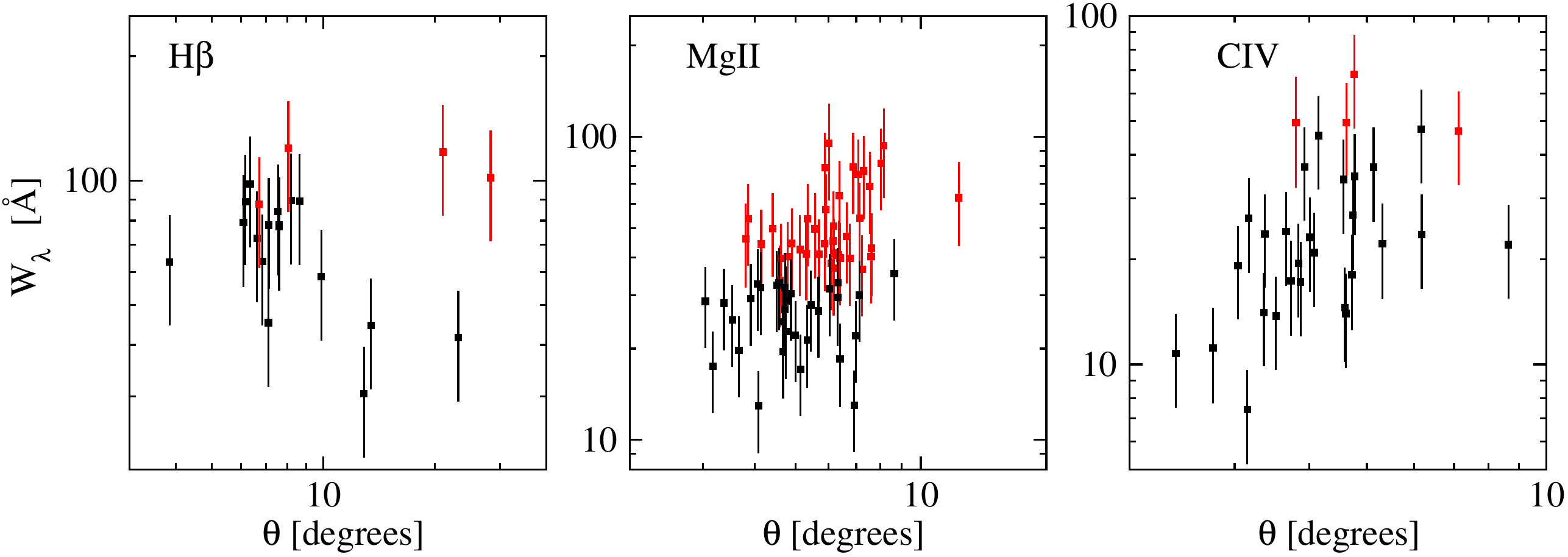}
\caption{Equivalent width vs. viewing angle of the jet. Red dots indicate sources with $NTD<1$. 
The Mg II emission line shows a significant correlation at confidence level $\geq$ 96\%. For the other lines there is no statistically significant correlation.}
\label{fig_ew_theta}
\end{figure*}

Boosting of the continuum emission at smaller viewing angles should lead to a decrease of emission line equivalent widths as a result 
of the increase of the line-continuum contrast. $W_\lambda$ and the jet viewing angle $\theta$ was compared in Figure~\ref{fig_ew_theta} for all three 
subsamples. A significant positive correlation was found between $W_\lambda$ and  $\theta$ for Mg II with $r=0.25$ and \textit{c.l.}=96.6\%, 
while the other lines do not show statistically significant results. It is worth noting that this correlation appears to be dominated by the points with $NTD<1$

In this work, the authors conclude that the equivalent width of Mg II is correlated with the jet viewing angle in the sense that increasing viewing angles produce 
larger values of $W_\lambda$, which is reasonable since larger viewing angles would mean less continuum boosting.

%


\section{Discussion and Conclusions}
\label{sec:disycon}

We investigate the Baldwin effect of 96 FSRQ for which spectroscopic data are available from \citet{torrealba2012,Torrealba14}. 
Our main results are the following: 

\begin{itemize}

\item 
$\;$We report that a significant Baldwin Effect was found in FSRQ, shown by significant anticorrelations 
(at the confidence level of $\geq\,95$\%) 
between equivalent widths of the H$\beta$,  Mg II and C IV emission lines and corresponding 
continuum luminosities.The slopes of the BE in FSRQ seem to be steeper than those in samples 
dominated by RQ quasars. Larger sampling of FSRQ is needed to confirm this result. 

\item
{$\;$The simulated toy model, shows that the difference we find in the slopes of the Baldwin Effect for RQ and FSRQ, 
cannot be explained by the addition of a non-thermal component to the continuum luminosity, nor by variability of the 
continuum components; which suggests that an extra emission line component is responsible for steepening the BE. }

\item
$\;$We found that roughly 80\% of FSRQ have a significant non-thermal contribution in optical/UV bands, i.e., 
\textit{NTD}$\,>\,$1, as was expected for AGN having relativistic jets. In particular, we reported that the bulk of 
UV emission is non-thermal and produced in the jet itself. The same evidence is corroborated for the optical 
continuum emission. 

\item
{$\;$We found values of $NTD<1$ for several FSRQ, and argue that this result  cannot be explained by a canonical 
BLR ionized only by the accretion disk, requiring the existence of an additional BLR component activated by the superluminal  jet 
at parsec scales down to the radio-core. Probably due to their larger sampling, this is shown by sources (57\%) 
with Mg II emission, and not in H$\beta$ and C IV samples.}

\item $\;$In both optical and UV emission \textit{NTD} increases at smaller viewing angles of the jet 
($\rm \theta\,\leq\,10^{\,\circ}$), which is in agreement with the prediction of the relativistic beaming theory.

\item $\;$A positive correlation was found between the equivalent width $W_\lambda$ and the viewing angle of the 
jet $\theta$ for Mg II, which is due to beaming of the continuum emission happening at small view angles 
that leads to a  decrease in $W_\lambda$ as a result of the continuum-line contrast.

\end{itemize}

It is well known that the SED of blazars are best described by beamed synchrotron emission from radio to X-ray frequencies, while 
the inverse Compton emission describes the SED of high frequency photons from the X-ray to TeV bands. The boosted synchrotron 
emission from the jet may dominate the low energy segment of the electromagnetic spectrum (radio to UV) in BL Lacs and FSRQ; while for radio galaxies 
(viewed at larger angles with respect to the jet) the emission from both the jet and accretion disk may significantly contribute to the total 
emission in optical, UV, and X-ray bands \citep[cf.][]{BR78,maraschi94,UP95}. The thermal emission from the accretion disk may dominate 
during the lower state of the jet activity (when the jet power is at minimum), and vice versa, the non-thermal emission of the jet would 
be dominant during the high states of the jet activity \citep[see Figure~4 in][]{arshakian08}.

This explains our findings presented earlier in section \ref{sec:NTD} above, that about 10\% to 25\% of the MOJAVE blazars 
from the H$\beta$ and C IV subsamples have a dominant thermal emission ($NTD<1$). 
We would expect that roughly the same percentage of FSRQ from the Mg II subsample are thermally dominated, 
in disagreement with our finding that about 50\% of sources have a thermal excess. For these quasars, the flux at 3000\,\AA\, 
coincides with the peak of the blue bump and the thermal luminosity at this frequency is higher than the luminosities at 1350\,\AA\, and 5100\,\AA\,
by about 0.1 \textit{dex}. This difference in luminosity is too small to reconcile the disagreement.
 Alternatively, the difference can be understood if there are two sources exciting the Mg II clouds, one is the thermal emission from the disk exciting 
the virialized Mg II clouds, and the other is the non-thermal jet emission which may excite both the virialized
 Mg II clouds and/or Mg II clouds outflowing along the jet \citep{Perez89,arshakian10_b,tavares10,tavares13}. 

In this scenario, the observed Mg II emission is reflecting the contribution from both the jet and accretion disk, and, hence, the predicted emission line luminosity cannot only be attributed to the accretion disk in radio-loud blazars. This inevitably leads to values of $NTD<1$ even for strongly non-thermal dominated sources. But the reason why only the Mg II line is affected by boosted jet emission remains unclear.

In the case of RQ AGN, the continuum emission radiation that ionizes the emitting gas regions comes from the thermal radiation of the accretion disk. While for FSRQ, the continuum emission also has a significant contribution from the non-thermal boosted radiation that arises from the relativistic jet \citep[e.g.,][]{D07,M08}. If the steepening of the Baldwin effect slope is inherent to FSRQ, this result could be directly related with the Doppler boosting of the continuum. If we could correct by Doppler boosting factor the continuum luminosity for all of our sources, we may find that the BE slope becomes flatter. But we need more data to do this analysis. Also a larger sample of FSRQ with optical and ultraviolet spectra is needed to confirm the BE slope steepening.


Our results listed above show that the non-thermal continuum of the jet contributes to the total continuum emission. 
Thus, the steep slopes of BE in blazars can be a signature of the contribution of the jet emission to the total continuum 
and line emission.

There is an observational evidence that optical flares and kinematics of the jet on sub-parsec scales are closely correlated: optical flare rises when superluminal component emerges into the jet. To explain the link between jet kinematics, optical continuum and emission line variability it was suggested the existence of the jet-excited BLR outflowing down the jet \citep{arshakian10_b, tavares10}.

\citet{tavares13} reported a flare-like event of the Mg II emission line during a $\gamma$-ray outburst in 
3C\,454.3. They found that the highest levels of the emission line flux coincide with a superluminal jet component 
traversing through the radio core, which was confirmed in consequent studies by \citet{isler13}. 
This is a direct observational evidence for a response of the broad emission lines to changes of the non-thermal 
continuum emission of the jet and, hence, the presence of the BLR material surrounding the radio core. The 
authors proposed an outflowing BLR which can arise from the accretion disk wind.  This possibility is supported and was previously 
suggested by \cite{Perez89}, and also for specific sources like 3C 273 \citep{paltani03} and 3C 454. 3 \citep{finke10,tavares13}.

From these previous findings combined with our results, we suggest the possibility that the jet emission greatly affects the gas 
that produces the emission lines and so, has an important contribution to the Baldwin Effect found in radio-loud compact AGN with superluminal jets. 
As a consequence, we can conclude that the relativistic plasma is tightly connected with the emitting line gas regions. 
The scenario we propose to explain the values of NTD$<1$ and the difference in slopes of the Baldwin Effect in RQ AGN and FSRQ (steeper in RQ) 
consists on a second component of the Broad Line Region that is related to the jet, probably in the form of an outflow. 
Further work is needed aiming to quantify the contribution of the jet emission to the total continuum emission, and as well, 
to find out the real distribution of the gas emitting region in blazar type AGN.

\section{Acknowledgments}
This work is based on observations acquired at the Observatorio Astron\'omico Nacional in the Sierra San Pedro M\'artir (OAN--SPM), 
Baja California, M\'exico, and at the Observatorio Astrof\'isico Guillermo Haro (OAGH), in Cananea, Sonora, M\'exico.
 This work is supported by CONACyT basic research grants 48484-F, 54480, and 151494 (Mexico). 
 V. P.-A. acknowledges sup- port from the CONACyT program for Ph.D. studies. 
 ICG acknowledges DGAPA (UNAM, Mexico) for a sabbatical scholarship and the Harvard-Smithsonian Center for Astrophysics for support as a visiting scholar.
TGA acknowledges support by DFG project number Os 177/2-1. L. \v C. P. is supported by the Ministry of Education and Science 
of R. Serbia through the project Astrophysical Spectroscopy of Extragalactic Objects (176001). The MOJAVE project is supported under 
National Science Foundation grant 0807860-AST and NASA-Fermi grant NNX08AV67G.

\clearpage


\begin{table}
\begin{center}
  \begin{threeparttable}
    \caption{. Parameters of weighted linear fitting for line and continuum luminosities ($\log\,L_{line}=A\,+B\,\log\,L_{c}$).}
	\label{tab:line_lum}
     \begin{tabular}{llccc}
        \toprule
        \multicolumn{1}{c}{\textbf{$L_c$}} &
	\multicolumn{1}{c}{\textbf{$L_{line}$}} &
	\multicolumn{1}{c}{\textbf{$A \pm \sigma_{\rm A}$}} &
	\multicolumn{1}{c}{\textbf{$B \pm \sigma_{\rm B}$}} &
	\multicolumn{1}{c}{\textbf{p}} \\
	\multicolumn{1}{c}{\textbf{(1)}} &
	\multicolumn{1}{c}{\textbf{(2)}} &
	\multicolumn{1}{c}{\textbf{(3)}} &
	\multicolumn{1}{c}{\textbf{(4)}} &
	\multicolumn{1}{c}{\textbf{(5)}} \\
	\midrule
	\multicolumn{5}{c}{RQ AGN} \\
	\midrule
	$ L_{5100}$ &        $ L_{\rm H\beta}$        &        $-$7.70$\,\pm\,$0.22        & 1.133$\,\pm\,$0.005        &        -----        \\
	$ L_{3000}$ &        $ L_{\rm Mg\,II}$        &        2.22$\,\pm\,$0.09        &        0.909$\,\pm\,$0.002        &        -----        \\
	$ L_{1350}$ &        $ L_{\rm C\,IV}$        &         4.42 $\,\pm\,$0.27        &        0.872$\,\pm\,$0.006        &       -----        \\
	\midrule
	\multicolumn{5}{c}{FSRQ} \\
	\midrule
	$L_{5100}$ 	&	$ L_{\rm H\beta}$	&	$-$1.32$\,\pm\,$7.58 &	0.988$\,\pm\,$0.166	 &	0.002	\\
	$L_{3000}$ 	&	$L_{\rm Mg\,II}$  	&	7.64$\,\pm\,$7.16	&	0.796$\,\pm\,$0.153	 &	0.012	\\
	$L_{1350}$ 	&	$L_{\rm C\,IV}$ 	&	8.32$\,\pm\,$7.04	&	0.788$\,\pm\,$0.148	 &	0.001	\\
        \bottomrule
     \end{tabular}
    \begin{tablenotes}
      \small
      \item The first part of the Table presents the relationships between line and continuum luminosities found by \citet{greene05} (H$\beta$) and \citet{shen11} (Mg II and C IV) for RQ samples. The second part shows our best weighted linear fit parameters for the $L_{line}$ vs. $L_{c}$, and to the equivalent width of emission line luminosity and the $L_{c}$ near each line in our blazar sample. Columns (1) and (2) are the continuum and emission line luminosity, respectively; Column (3) is the intercept and its error; Column (4) is the linear fit slope and its error; Column (5) is the statistical probability of the weighted linear fit, p$\,\leq\,0.05$ means that the linear correlation is statistically significant at a $c.l.\geq\,95\%$.
    \end{tablenotes}
  \end{threeparttable}
\end{center}
\end{table}


\begin{table}
\begin{center}
  \begin{threeparttable}
	\caption{. Parameters of weighted linear fitting for the Baldwin Effect ($\log\,W_{\lambda}=\alpha\,+\beta\,\log\,L_{c}$).}
    \label{tab:line_c_EW}
     \begin{tabular}{llccc}
        \toprule
         \multicolumn{1}{c}{\textbf{$L_c$}} &
	\multicolumn{1}{c}{\textbf{$W_\lambda(line)$}} &
	\multicolumn{1}{c}{\textbf{$\alpha \pm \sigma_{\alpha}$}} &
	\multicolumn{1}{c}{\textbf{$\beta \pm \sigma_{\beta}$}} &
	\multicolumn{1}{c}{\textbf{p}} \\
	\multicolumn{1}{c}{\textbf{(1)}} &
	\multicolumn{1}{c}{\textbf{(2)}} &
	\multicolumn{1}{c}{\textbf{(3)}} &
	\multicolumn{1}{c}{\textbf{(4)}} &
	\multicolumn{1}{c}{\textbf{(5)}} \\
	\midrule
	\multicolumn{5}{c}{RQ AGN } \\
	\midrule
	$ L_{5100}$ &        $W_\lambda$(H$\beta$)           & $-$4.864$\,\pm\,$0.708         & 0.149$\,\pm\,$0.013        &        --        \\
	$ L_{3000}$ &        $W_\lambda$(Mg II)        & 5.798$\,\pm\,$0.099             &  $-$0.093$\,\pm\,$0.002   &        --        \\
	$ L_{1350}$ &        $W_\lambda$(C IV)          & 7.718$\,\pm\,$0.154              &  $-$0.131$\,\pm\,$0.003   &        --        \\
	\midrule
	\multicolumn{5}{c}{FSRQ} \\
	\midrule
	$L_{5100}$ 	&	 $W_\lambda$(H$\beta$)	     &	2.41$\,\pm\,$2.78	&	$-$0.012$\,\pm\,$0.061 	&	0.013 	\\
	$L_{3000}$ 	&	 $W_\lambda$(Mg II) 	&	13.38$\,\pm\,$4.97	&	$-$0.253$\,\pm\,$0.107 	&	0.011 	\\
	$L_{1350}$ 	&	 $W_\lambda$(C IV)  	&	11.66$\,\pm\,$4.02	&	$-$0.216$\,\pm\,$0.085 	&	0.010	     \\
        \bottomrule
     \end{tabular}
    \begin{tablenotes}
      \small
      \item The first part of the Table presents the relationships between $W_{\lambda}$ and continuum luminosities simulated from line-luminosity relations in \citet{greene05} (H$\beta$) and \citet{shen11} (Mg II and C IV) for RQ AGN. The second part shows, for the FSRQ sample, our best weighted linear fit parameters for the relation $W_{\lambda}$ vs. $L_{c}$, for the equivalent width of each emission line and the corresponding $L_{c}$. Columns (1) and (2) are the continuum and equivalent width, respectively; Column (3) is the intercept and its error; Column (4) is the linear fit slope and its error; Column (5) is the statistical probability of the weighted linear fit, p$\,\leq\,0.05$ means that the linear correlation is statistically significant at a $c.l.\geq\,95\%$.
    \end{tablenotes}
  \end{threeparttable}
\end{center}
\end{table}

\clearpage


\bibliographystyle{frontiersinSCNS_ENG_HUMS.bst}  
\bibliography{Baldwin_Effect_references} 

\end{document}